\RequirePackage{amsmath} 
\documentclass[12pt]{article}
\usepackage{a4wide}
\usepackage[utf8]{inputenc}
\usepackage[title]{appendix}
\usepackage[normalem]{ulem}
\usepackage[]{algorithm2e}
\SetKwInput{KwData}{Input}
\SetKwInput{KwResult}{Output}
\usepackage{authblk}
\usepackage{tikz}
\usetikzlibrary{plotmarks}
\usepackage{pgfplots,pgfplotstable}
\usepackage{comment}

\usepackage{hyperref}

\usepackage{arydshln}

\usepackage{amsfonts,amssymb,amsthm,mathtools}
\usepackage{color} 

\newcommand{\ff}[1]{\mathbb{F}_{#1}} 

\newcommand{\Fq}{\ff{q}}
\newcommand{\Fqm}{\ff{q^m}}
\newcommand{\NN}{\mathbb{N}} 

\newcommand{\any}{*} 
\newcommand{\rank}[1]{\operatorname{Rank}\mathchoice{\left(#1\right)}{(#1)}{(#1)}{(#1)}} 
\newcommand{\rw}[1]{\left| #1 \right|_{\text{RANK}}}
\DeclareMathOperator{\Mat}{Mat}

\DeclareMathOperator{\MaxMinors}{MaxMinors}

\DeclareMathOperator{\LT}{LT}
\newcommand{\minor}[2]{\left\vert #1 \right\vert_{#2}}

\newcommand{\matRing}[3]{#1^{#2 \times #3}} 
\newcommand{\mat}[1]{\boldsymbol{#1}} 
\newcommand{\trsp}[1]{#1^\mathsf{T}} 
\newcommand{\cv}{\mat{c}}
\newcommand{\ev}{\mat{e}}
\newcommand{\vv}{\mat{v}}
\newcommand{\xv}{\mat{x}}

\newcommand{\sv}{\mat{s}}
\newcommand{\yv}{\mat{y}}
\newcommand{\zerom}{\mat{0}}

\newcommand{\Cm}{\mat{C}}

\newcommand{\Em}{\mat{E}}

\newcommand{\Hm}{\mat{H}}
\renewcommand{\Im}{\mat{I}}
\newcommand{\Lm}{\mat{L}}
\newcommand{\Mm}{\mat{M}}

\newcommand{\Rm}{\mat{R}}
\newcommand{\Sm}{\mat{S}}

\newcommand{\Um}{\mat{U}}

\newcommand{\Dc}{{\mathcal D}}
\newcommand{\Cc}{{\mathcal C}}

\newcommand{\eqdef}{:=} 

\newcommand{\E}{\mathrm{E}}
\newcommand{\Var}{\mathrm{Var}}

\usepackage[capitalize]{cleveref}

\makeatletter
\newcommand\bibalias[2]{%
	\@namedef{bibali@#1}{#2}%
}

\newtoks\biba@toks
\newcommand\acite[2][]{%
	\biba@toks{\cite#1}%
	\def\biba@comma{}%
	\def\biba@all{}%
	\@for\biba@one:=#2\do{%
		\@ifundefined{bibali@\biba@one}{%
			\edef\biba@all{\biba@all\biba@comma\biba@one}%
		}{%
			\PackageInfo{bibalias}{%
				Replacing citation `\biba@one' with `\@nameuse{bibali@\biba@one}'
			}%
			\edef\biba@all{\biba@all\biba@comma\@nameuse{bibali@\biba@one}}%
		}%
		\def\biba@comma{,}%
	}%
	\edef\biba@tmp{\the\biba@toks{\biba@all}}%
	\biba@tmp
}
\makeatother

\bibalias{Ouroboros-R}{AABBBDGHZ17}
\bibalias{RQC}{AABBBDGZ17}
\bibalias{RQC2}{AABBBDGZCH19}
\bibalias{ROLLO}{ABDGHRTZABBBO19}
\bibalias{LAKE}{ABDGHRTZ17}
\bibalias{LOCKER}{ABDGHRTZ17a}
\bibalias{RankSign}{AGHRZ17}

\DeclareMathOperator{\Unfold}{{\bf UnFold}}

\newtheorem{assumption}{Assumption}
\newtheorem{problem}{Problem}

\newtheorem{prop}{Proposition}

\newtheorem{rem}{Remark}
\newtheorem{defi}{Definition}
\newtheorem{modeling}{Modeling}
\newtheorem{lemma}{Lemma}
\newtheorem{theorem}{Theorem}

\begin{document}
	
\title{An algebraic approach to the Rank Support Learning problem}
\author[1,3]{Magali Bardet \thanks{magali.bardet@univ-rouen.fr}}
\author[2,3]{Pierre Briaud \thanks{pierre.briaud@inria.fr}}
\affil[1]{LITIS, University of Rouen Normandie}
\affil[2]{Sorbonne Universit\'es, UPMC Univ Paris 06}
\affil[3]{Inria, Team COSMIQ,
  2 rue Simone Iff, CS 42112,
  75589 Paris Cedex 12, France}
\date{}

\maketitle
	
\begin{abstract}
  Rank-metric code-based cryptography relies on the hardness of
  decoding a random linear code in the rank metric. The Rank Support
  Learning problem (RSL) is a variant where an attacker has access to
  $N$ decoding instances whose errors have the same support and wants
  to solve one of them. This problem is for instance used in the
  Durandal signature scheme \cite{ABGHZ19}. In this paper, we propose
  an algebraic attack on RSL which clearly outperforms the previous
  attacks to solve this problem. We build upon \cite{BBCGPSTV19},
  where similar techniques are used to solve MinRank and RD. However,
  our analysis is simpler and overall our attack relies on very
  elementary assumptions compared to standard Gröbner bases
  attacks. In particular, our results show that key recovery attacks
  on Durandal are more efficient than was previously thought.
		
  \paragraph{keywords} {Post-quantum cryptography - rank metric code-based
    cryptography - algebraic attack.}
\end{abstract}
	
\section{Introduction.}
\label{sec:inro}

\paragraph{Rank metric code-based cryptography.}
In the last decade, rank metric code-based cryptography has proved to be a powerful alternative to traditional code-based cryptography based on the Hamming metric. Compared to the situation in the Hamming metric, a few families of codes with an efficient decoding algorithm were considered in rank-based cryptography. Starting with the original GPT cryptosystem \cite{GPT91}, a first trend was to rely on Gabidulin codes. However, their algebraic structure was successfully exploited by the Overbeck attack \cite{O05} and variants. More recent proposals \acite{GMRZ13,GRSZ14,LAKE,LOCKER} inspired by the NTRU cryptosystem \cite{HPS98} were based on LRPC codes. These schemes can be viewed as the rank metric analogue of the MDPC cryptosystem in the Hamming metric \cite{MTSB12}, where the trapdoor is given by a small weight dual matrix which allows efficient decoding. 

The cryptosystems submitted to the NIST post-quantum Standardization Process  \acite{ABDGHRTZABBBO19,AABBBDGZCH19} were of this kind. They have not passed the second round of this competition, but the NIST still encourages further research on rank-based cryptography. First, they offer an interesting gain in terms of public-key size due to the underlying algebraic structure. Also, this type of cryptography is not restricted to the abovementioned encryption schemes, as shown by a proposal for signature \cite{ABGHZ19} and even the $\mathsf{IBE}$ scheme from \cite{GHPT17a}, and more progress might be made in that direction.

\paragraph{Decoding problems in rank metric.} Codes used in rank metric cryptography are linear codes over an extension field $\ff{q^m}$ of degree $m$ of $\ff{q}$. An $\ff{q^m}$-linear code of length $n$ is an $\Fqm$-linear subspace of $\Fqm^n$, but codewords can also be viewed as matrices in $\ff{q}^{m \times n}$. Indeed, if $(\beta_1,\dots,\beta_m)$ is an $\ff{q}$-basis of $\ff{q^m}$, the word $\xv = (x_1,\ldots,x_n) \in \ff{q^m}^{n}$ corresponds to the matrix $\Mat(\xv) = (X_{ij})_{i,j} \in \ff{q}^{m \times n}$, where $x_j = \beta_1X_{1j} + \dots + \beta_m X_{mj}$ for $j \in \{1..n\}$. The weight of $\xv$ is then defined by using the underlying rank metric on $\ff{q}^{m \times n}$, namely
$\rw{\xv} := \rank{\Mat{(\xv)}}$, and it is also equal to the dimension of the \emph{support} $\text{Supp}(\xv) := \langle x_1,\dots,x_n \rangle_{\ff{q}}$. Similarly to the Hamming metric, the main source of computational hardness for rank-based cryptosystems is a decoding problem. It is the decoding problem in rank metric restricted to $\Fqm$-linear codes, namely
\begin{problem} (Rank Decoding problem (RD))\\
	\indent	\emph{Input}: an $\Fqm$-basis $(\cv_1,\dots,\cv_k)$ of a subspace $\Cc$ of
	$\Fqm^n$, an integer $r \in \NN$, and a vector $\yv \in \Fqm^n$ such that $\rw{\yv-\cv} \leq r$ for some $\cv \in \Cc$. \\
	\indent	\emph{Output}: $\cv \in \Cc$ and an \emph{error} $\ev \in \Fqm^n$  such that $\yv=\cv+\ev$
	and $\rw{\ev} \leq r$.
\end{problem}
We also adopt the \emph{syndrome} formulation: given $\sv \in \ff{q^m}^{n-k}$ and $\mat{H}\in \ff{q^m}^{(n-k) \times n}$ a parity-check matrix of the code, find $\ev \in \ff{q^m}^n$ such that $\mat{H}\trsp{\ev} = \trsp{\sv}$ and $\rw{\ev} \leq r$. Without this restriction to $\Fqm$-linear codes, the decoding of arbritrary codes in rank metric is also worthy of interest. It is equivalent to the following MinRank problem, as explained in \cite{FLP08}. 
\begin{problem} (MinRank problem)\\
	\indent\emph{Input}: an integer $r \in \NN$ and $K$ 
	matrices $\mat{M}_1,\dots,\mat{M}_{K} \in \matRing{\ff{q}}{m}{n}$.\\
	\indent\emph{Output}:  field elements $x_1,x_2,\dots,x_{K} \in \ff{q}$, not all zero, such that
	\begin{equation*}
		\text{Rank}\left(\sum_{i=1}^{K}x_i\Mm_i\right) \leq r.
	\end{equation*}
\end{problem}
This problem was originally defined and proven NP-complete in \cite{BFS99}, and it is now ubiquitous in multivariate cryptography. However, the RD problem is not known to be NP-complete, and there only exists a randomized reduction to the decoding in the Hamming metric, which is NP-complete \cite{GZ14}. Still, this problem is believed to be hard and the best attacks have exponential complexity. The first attacks against RD were of combinatorial nature \cite{GRS13}\cite{HT15a}\cite{AGHT18}, but the recent developments in \cite{BBBGNRT19}\cite{BBCGPSTV19} tend to show that algebraic methods now perform better, even for small values of $q$.  
\paragraph{The RSL problem.} The Rank Support Learning problem is a generalization of the RD problem to several syndromes which correspond to errors with the same support.
\begin{problem} (Rank Support Learning (RSL))\\
	\indent	\emph{Input}: $(\mat{H},\mat{H\trsp{E}})$, where $\mat{H} \in \ff{q^m}^{(n-k) \times n}$ is full-rank and $\mat{E} \in \ff{q^m}^{N \times n}$ has all its entries lying in a subspace $\mathcal{V} \subset \ff{q^m}$ of dimension $r$ for some $ r\in \mathbb{N}$. \\
	\indent	\emph{Output}: The secret subspace $\mathcal{V}$.
\end{problem}
In other words, an RSL instance of parameters $(m,n,k,r,N)$ consists of $N$ RD instances $\mat{H}\trsp{\ev_i} = \trsp{\sv_i}$ with common support $\mathcal{V}$ of dimension $r$ for $i \in \{1..N\}$. The RSL problem can be seen as the rank metric analogue of the Support-Learning problem in the Hamming metric, which had already been used for cryptographic purposes \cite{KKS97}\cite{OT11}. However, the RSL problem turns out to be much more versatile for the future of rank-based cryptography. It was introduced in \cite{GHPT17a} to build an $\mathsf{IBE}$ scheme, broken lately in \cite{DT18b}. More importantly, this problem is at the heart of the security of the Durandal signature scheme \cite{ABGHZ19}, and solving the underlying RSL instance leads to a key recovery attack. It is readily apparent that the difficulty of RSL decreases when the number of RD instances grows. On the one hand, the RSL problem is equivalent to RD when $N=1$, and therefore the best known attacks are exponential in the parameters $(m,n,k,r)$. On the other hand, difficult instances must satisfy $N < kr$, as explained in \cite{DT18b}. So far, the only attempt to solve RSL for all values of $N$ was the combinatorial algorithm from the original RSL paper \cite{GHPT17a}, which leaves room for improvement. 

\paragraph{Contribution.} Our contribution is an algebraic attack on the RSL problem and therefore a key recovery attack on Durandal. To the best of our knowledge, it is the only attack on RSL when $N<kr$ which is not an attack on RD since \cite{GHPT17a}. Note that the Durandal current parameter sets were already broken by the algebraic attacks from \cite{BBCGPSTV19} and have not been updated since then. Therefore, we propose new parameters in order to avoid these attacks as well as the other known attacks on Durandal (see Section \ref{sec:tables}), and in Table \ref{tab:rsl_short} we compare our attack to the best existing attack on RSL for these parameters. 

\begin{table}
  \centering
	\caption{Complexity of our attack on parameters $(m,n,k,r,N)$ corresponding to Durandal parameter sets given in Table \ref{tab:rsl_long}. ``Best RD" refers to the RD attack from \cite{BBCGPSTV19} which is the best RSL attack so far. The last two columns correspond to our attack for the two values of $N$ considered in Durandal. An underlined value is an improvement upon the RD attack.}
	\begin{tabular}{|*{4}{c|}}
		\hline
		$(m,n,k,r)$ &  Best RD &  $N=k(r-2)$ & $N=k(r-1)$  \\\hline
		
		$(277,358,179,7)$ & 130 & \underline{125} & \underline{126}  \\ 
		\hline
		$(281,242,121,8)$ & 159  & 170 & \underline{128}  \\
		\hline
		$(293,254,127,8)$ & 152  & 172 & \underline{125} \\ 
		\hline
		$(307,274,137,9)$  & 251 & \underline{187} & \underline{159} \\
	\hline
	
	\end{tabular}\label{tab:rsl_short}
\end{table}

Our attack is very often more efficient than the best current RSL attack, especially for a large number of errors. The improvement is also more significant for larger values of parameters (see for instance the last row in Table \ref{tab:rsl_short} compared to the other ones). Regarding the security of Durandal, our work greatly improves upon previous key recovery attacks and therefore it will have to be taken into account when selecting future parameters for the scheme.

The original attack from \cite{GHPT17a} is a combinatorial algorithm to look for elements of low weight in a code $\Cc_{aug} := \Cc + \langle \ev_1,\dots,\ev_N \rangle_{\ff{q}}$ of typical dimension $km+N$ which contains many such codewords. Our approach is to attack the very same code $\Cc_{aug}$ but by using algebraic techniques. A direct adaptation would be to consider a MinRank instance with $km + N$ matrices in $\mathbb{F}_{q}^{m \times n}$ which represent an $\ff{q}$-basis of  $\Cc_{aug}$. However, the region of parameters used in rank-based cryptography is typically $m = \Theta(n)$ and $k = \Theta(n)$, so that the number of matrices is $\Theta(n^2)$ due to the term $km$. This makes the cost of this approach too high to be relevant. Therefore, we propose a bilinear modeling of the problem with only $N$ matrices in $\mathbb{F}_{q}^{m \times n}$ instead of $km + N$.
The way this system is obtained is very reminiscent of the work of \cite{BBCGPSTV19} to attack MinRank and RD. First, it consists of the set of all maximal minors of a matrix of linear forms over $\ff{q^m}$ which are then ``descended" over $\ff{q}$ as in the MaxMinors modeling. Second, we adopt a similar $\lambda$-XL type strategy by multiplying the initial equations by monomials in only one of the two blocks of variables. The system is then solved by linearization at some bi-degree $(b,1)$. To determine precisely this degree, we have to carefully count the number of independent equations at each bi-degree. In the case of MinRank, Bardet \emph{et al.} are able to construct explicit linear relations between the augmented equations and they argue that the rest of the equations are linearly independent \cite[Heuristic $2$, p.19]{BBCGPSTV19}. Their counting is valid whenever $b<r+2$, where $r$ is the target rank in MinRank. However, our analysis will be much tighter. Indeed, up to a minor assumption on the RSL instance that can be easily checked by linear algebra on the syndromes, we can construct an explicit basis for the rowspace of the Macaulay matrix at each bi-degree $(b,1)$, and we do no longer have a limitation on the value of $b$ apart from $b < q$ {($q\ne 2$)}.\\
Also, we do not restrict ourselves to the words of lowest weight in $\Cc_{aug}$ as in \cite{GHPT17a}. The reason is that decreasing the target weight $r$ as much as possible is always advantageous for the combinatorial attacks, but not necessarily for the algebraic attacks. Indeed, decreasing $r$ will cause to decrease both the number of equations and variables in the system, but the ratio between the two might become defavorable.

\paragraph{Notation.}
\label{sec:prelim}

For $a,b$ integers such that $a \leq b$, we denote by $\{a..b\}$ the set of integers from $a$ to $b$. The notation $\# I$ stands for the cardinality of the finite set of integers $I$, and for $a$ an integer, $I+a$ stands for the set $\lbrace i+a : i \in I\rbrace$. Also, we denote by $Pos(i,I)$ the position of the integer $i$ in the ordered set $I$. \\ The space of matrices of size $m \times n$ over a field $\mathbb{K}$ is denoted by $\mathbb{K}^{m \times n}$. Matrices and vectors are denoted by bold lowercase letters $(\Mm,\vv)$. For $I\subset\{1..n\}$ and $J\subset\{1..m\}$, we use the notation $\mat M_{I,J}$ for the submatrix of $\mat M$ formed by its rows
(resp.~columns) with indexes in $I$ (resp.~$J$). We adopt the shorthand notation $\mat M_{\any,J} = \mat{M}_{\{1..m\},J}$
and $\mat M_{I,\any} = \mat{M}_{I,\{1..n\}}$, where \(\mat M\) has size $m \times n$. 

\section{Durandal and the RSL problem.}
\label{sec:durandal}
		
Assessing the hardness of RSL is needed to evaluate the security of the Durandal signature scheme \cite{ABGHZ19}. This scheme is based on the Lyubashevsky framework adapted to the rank metric setting. For a $128$-bit security level, the original parameters offer a signature size of less than 4kB and a public key of size less than 20kB. 

\subsection{Key pair in Durandal.}\label{sec:key_pair}

Durandal is an authentication protocol turned into a signature thanks to the Fiat-Shamir transform. For the purposes of this paper, we simply describe the key pair and we refer the reader to \cite[§3]{ABGHZ19} for a full presentation of this protocol. First, the secret key consists of a couple of matrices $(\Em_1,\Em_2) \in \ff{q^m}^{lk \times n} \times \ff{q^m}^{l'k \times n}$ whose entries lie in a subspace $\mathcal{V}$ of dimension $r$. The public key is $(\Hm,\Sm_1 \vert \Sm_2)$ such that $\Hm \in \ff{q^m}^{(n-k)\times n}$ is a random full-rank ideal matrix \footnote{See for instance \cite[§2.2, Definition $8$]{ABGHZ19} for a definition.}, $\Sm_1 = \Hm\trsp{\Em_1} \in \ff{q^m}^{(n-k)\times lk}$ and $\Sm_2 = \Hm{\Em_2}^{\mathsf{T}} \in \ff{q^m}^{(n-k)\times l'k}$, where $\vert$ denotes matrix concatenation. It is readily verified that the couple $(\Hm,\Sm_1 \vert \Sm_2)$ is an instance of RSL with parameters $(m,n,k,r)$ and $N = lk + l'k$, and that solving this instance leads to a key-recovery attack. However, it is not a \emph{random} instance. Indeed, the matrix $\Sm_1$ (resp. $\Sm_2$) can be reconstructed from only $l$ (resp. $l'$) of its columns due to the ideal structure of $\Hm$. However, we have not been able to exploit this fact, and since this extra structure is only used for efficiency, we assume without loss of generality that we attack a random RSL instance. 

\subsection{Previous cryptanalysis on RSL.}

The security of Durandal relies on the hardness of RD and RSL, as well as of the $\text{PSSI}^{+}$ problem, which is an \emph{ad hoc} assumption \cite[Problem $5$]{ABGHZ19}. In this section, we describe the prior work on RSL which was considered to design the parameters. 

\paragraph{Attacks for large $N$.} First, the RSL problem becomes easy when $N \geq nr$ and a polynomial attack is detailed in \cite[§4.2, p.14]{GHPT17a}. This linear algebra argument is not really specific to the rank metric in the sense that it can be applied to the very same problem in the Hamming metric. A more powerful attack is given in \cite{DT18b} and suggests that secure instances of RSL must satisfy the stronger condition $N < kr$. The idea is that when $N \geq kr$, the public $\ff{q}$-linear code
\begin{equation*}
	\mathcal{C}_{synd} := \left\lbrace \xv\mat{E\trsp{H}},~\xv \in \ff{q}^N \right\rbrace 
\end{equation*}
is such that $\dim_{\ff{q}}{(\mathcal{C}_{synd} \cap \mathcal{V}^{n-k})} \geq N - kr$, and therefore there exist at least $q^{N - kr}$ words of weight $r$ in $\mathcal{C}_{synd}$. The authors propose a bilinear modeling to recover one of these codewords and due to the high number of solutions, many variables can be eliminated from the system. The attack is efficient because the Gröbner basis techniques on this system are expected to take subexponential time. However, it seems difficult to adapt the argument of \cite{DT18b} for $N$ even slightly below $kr$, because the intersection $\mathcal{C}_{synd} \cap \mathcal{V}^{n-k}$ will be trivial. Therefore, the Durandal parameter sets are chosen such that $N = (k-2)r$ or $N = (k-1)r$ and the complexity analysis is based on the original attack from \cite{GHPT17a}. 

\paragraph{Solving RSL when $N < kr$.}
A naive way to solve RSL when $N < kr$ is to attack one of the $N$ RD instances. Following \cite{OJ02}, the strategy is to look for words of weight $\leq r$ in an augmented $\ff{q^m}$-linear code of the form $\Cc_{\ev} = \Cc \oplus \langle \ev \rangle$. To tackle several errors, note that adding $\ev_1, \dots, \ev_N$ to the code $\Cc$ in an $\ff{q^m}$-linear manner will lead to a deadlock, because the augmented code quickly covers the whole space $\ff{q^m}^{n}$. Therefore, the authors of \cite{GHPT17a} consider a code containing all the errors but which is simply $\ff{q}$-linear. Let $W_{U}  \subset \ff{q^m}^{n-k}$ be the $\ff{q}$-linear space generated by the $\trsp{\sv_i}$ for $i \in \{1..N\}$ and let $\Cc_{aug}$ be the $\ff{q}$-linear code defined by 
\begin{equation*}
	\Cc_{aug} := \left \lbrace \xv \in \ff{q^m}^n,~\mat{H}\xv \in W_{U}  \right \rbrace.
\end{equation*}
We clearly have $\Cc' \subset \Cc_{aug}$, where $\Cc'$ is defined by $\Cc' := \left \langle \trsp{\ev_1}, \dots, \trsp{\ev_N} \right \rangle_{\ff{q}}$. Codewords in $\Cc'$ all have weight $\leq r$, and therefore the code $\Cc_{aug}$ typically contains $q^N$ words of this weight. It also contains the public $\ff{q^m}$-linear code $\Cc :=  \left \lbrace \xv \in \ff{q^m}^n,~\mat{H}\xv=0 \right \rbrace$. We have $\dim_{\ff{q}}\Cc_{aug} \leq \dim_{\ff{q}}\Cc + \dim_{\ff{q}}W_U \leq km + N$. In general, this inequality is an equality and we will make make this assumption from now on. In particular, it implies that the errors $\ev_1, \dots, \ev_N$ are linearly independent over $\ff{q}$. The authors propose a combinatorial algorithm \cite[§4.3, Algorithm 1]{GHPT17a} to look for low weight codewords in $\Cc_{aug}$. Their attack greatly benefits from the fact that there are many words of weight $r$ in $\Cc'$ and, a fortiori, in $\Cc_{aug}$. Indeed, the algorithm will still succeed by targeting a word of weight equal to the minimum distance of $\Cc'$.
This leads to the complexity claimed in \cite[Theorem $2$]{GHPT17a}, which is equal to $q^{min(e_{-},e_{+})}$, where {$K = km + N$ and}
 \begin{align*}
 	e_{-}  & = \left(w - \left\lfloor \frac{N}{n} \right\rfloor \right)\left(\left\lfloor \frac{K}{n} \right\rfloor - \left\lfloor \frac{N}{n} \right\rfloor \right) \\ 
 	 e_{+}  & = \left(w - \left\lfloor \frac{N}{n} \right\rfloor - 1\right)\left(\left\lfloor \frac{K}{n} \right\rfloor - \left\lfloor \frac{N}{n} \right\rfloor - 1\right) + n\left(\left\lfloor \frac{K}{n} \right\rfloor - \left\lfloor \frac{N}{n} \right\rfloor - 1\right).
 \end{align*}

\section{The RSL-Minors modeling.}
\label{sec:attack1}
In this section, we introduce the algebraic modeling that we use to solve the RSL problem and we propose two ways to restrict the number of solutions, so that the final system has roughly one solution.

\subsection{The basic modeling.}\label{sec:the_modeling}

Our system is obtained as follows. First, a public basis of the code $\Cc_{aug}$ can be obtained by considering an $\ff{q^m}$-basis of $\Cc$ (\emph{i.e.} a full-rank generator matrix $\mat{G}\in\ff{q^m}^{k \times n}$) together with elements $\yv_i \in \ff{q^m}^n$ such that $\yv_i \trsp{\mat{H}} = \sv_i$ for $i \in \{1..N\}$. A word of weight $w \leq r$ in $\Cc_{aug}$ is then written as
\begin{equation*}
\ev := x\mat{G} + \sum_{i=1}^{N}\lambda_i\yv_i := (\beta_1,\beta_2,\dots,\beta_m)\mat{C}\mat{R},
\end{equation*} 
where the quantities $x \in \ff{q^m}^k$, $\lambda_i \in \ff{q}$ for $i \in \{1..N\}$, $\mat{C} \in \ff{q}^{m \times w}$ and $\mat{R} \in \ff{q}^{w \times n}$ are  unknowns \footnote{We adopt this notation because the matrix $\Rm$ (resp. $\Cm$) represents a basis of the Rowspace (resp. Column space) of $\Mat{(\ev)}$.}. Since $\mat{G}\trsp{\mat{H}} = 0$, variables can be removed by multiplying to the right by $\trsp{\mat{H}}$, and one obtains
\begin{equation*}\label{sys:smsys}
	\sum_{i=1}^{N} \lambda_{i}\sv_{i} = (\beta_1,\dots,\beta_m)\mat{C}\mat{R}\trsp{\mat{H}}. 
\end{equation*}
The vector $\sum_{i=1}^{N}\lambda_{i}\sv_{i}$ is a linear combination over $\mathbb{F}_{q^m}$ of the rows of $\mat{R}\trsp{\mat{H}}$. This means that the following matrix   
\[
\Delta_{\Hm} :=\begin{pmatrix}
	\sum_{i=1}^{N} \lambda_{i}\sv_{i} \\
	\mat{R}\trsp{\mat{H}}
\end{pmatrix} = \begin{pmatrix}
	\sum_{i=1}^{N} \lambda_{i}\yv_{i} \\
	\mat{R}
\end{pmatrix}\trsp{\mat{H}}  \in \mathbb{F}_{q^m}^{(w+1) \times (n-k)}
\] 
has rank at most $w$. Finally, equations are obtained by canceling all the maximal minors of $\Delta_{\Hm}$. They are labelled by all the subsets $J \subset \lbrace 1..n-k \rbrace$ of size $w+1$.
	\begin{align}
	\mathcal{F} & =  \left\{f = 0 \Big| f \in 
	\MaxMinors(\Delta_{\Hm}) \right\} \label{eq:RSL_eq_fqm} \\
	& = \left\{\vert\Delta_{\Hm}\vert_{*,J} = 0 \Big| J \subset \lbrace 1..n-k \rbrace,~\# J=w+1\right\}. \nonumber
\end{align}

The following Lemma \ref{lemma:QJ} shows that the equations are bilinear in the $\lambda_i$ and in the $r_T$ variables, which are the maximal minors of $\Rm$. 
\begin{lemma}
	\label{lemma:QJ}
	Let $J \subset \lbrace 1..n-k \rbrace$ such that $\# J = w+1$. We have
	\begin{align*}
		Q_J & \eqdef \minor{\Delta_{\Hm}}{\any,J} 
		= \sum_{i=1}^{N} \lambda_{i} \sum_{\substack{T \subset
				\lbrace 1..n \rbrace \\ \#T = w}} r_{T} \sum_{t \notin
			T} y_{i,t}(-1)^{1+Pos(t,T \cup \lbrace t \rbrace)}\left\vert
		\Hm \right \vert_{J,T \cup \lbrace t \rbrace},
		\label{eq:QJ}
	\end{align*}
where $r_T = \left\vert\Rm\right\vert_{\any,T},~T \subset \{1..n\},~\#T=w$. Without loss of generality, we assume that $\Hm_{\any,\lbrace k+1..n\rbrace} = \Im_{n-k}$, so that $Q_J$ contains $N\binom{k+1+w}{w}$ monomials.
\end{lemma}
The proof can be found in Appendix~\ref{sec:appendixA}. Since the equations have coefficients in $\ff{q^m}$ and solutions $\lambda_i$, $r_T$ are searched in $\ff{q}$, we \emph{unfold} the system over $\ff{q}$. It consists in expanding each equation $f$ over $\ff{q^m}$ as $m$ equations $[f]_{j}$ over $\ff{q}$ for $j \in \{1..m\}$ which represent the ``coordinates" of $f$ in an $\ff{q}$-basis of $\ff{q^m}$.
\begin{modeling}[RSL Minors modeling]\label{mod:RSL_minors_modeling}
	We consider the system over $\ff{q}$ obtained by unfolding the system \eqref{eq:RSL_eq_fqm}:
	\begin{equation}\label{eq:RSL_eq}
		\Unfold(\mathcal{F}) = \Unfold\left(\left\{f = 0 \Big| f \in 
		\MaxMinors(\Delta_{\Hm}) \right\}\right).
	\end{equation}
	This system contains:
	\begin{itemize}
		\item $m \binom{n-k}{w+1}$ bilinear equations with coefficients in $\Fq$,
		\item  $N+\binom{n}{w}$ unknowns: $\mathbf{\lambda}=(\lambda_1,\cdots,\lambda_{N})$ and the $r_T$'s, where $r_T = \left\vert\Rm\right\vert_{\any,T}$ for $T \subset \{1..n\},~\#T=w$.
	\end{itemize}
	We search for solutions $\lambda_i, r_T$'s in $\Fq$.
\end{modeling}
We now describe two ways to restrict the number of solutions to the RSL Minors system. Note that the weight $w \leq r$ in Modeling \ref{mod:RSL_minors_modeling} is not set to a precise value, and contrary to \cite{GHPT17a}, we will not necessarily target the words of lowest weight in $\Cc_{aug}$. Actually, we prefer to attack codes obtained by shortening $\Cc_{aug}$.
\begin{defi}[Shortening a matrix code]
	Let $\Cc_{mat}\subset\ff{q}^{m \times n}$ be a matrix code of parameters $[m \times n,K]_{q}$ and $I \subset \{1..n\}$. The shortening $\mathcal{S}_{I}(\Cc_{mat}) \subset \ff{q}^{m \times (n - \#I)}$ of $\Cc_{mat}$ is the $[m \times (n - \#I),K' \geq K - m\#I]_{q}$-code defined as follows:
	\begin{equation*}
		\mathcal{S}_{I}(\Cc_{mat}) := \left\lbrace \mat{R}_{*,\{1..n\}\setminus I}~|~ \mat{R} \in \Cc_{mat},~ \mat{R}_{*,I} = 0_{*,I} \right\rbrace.
	\end{equation*}
	Moreover, when the code $\Cc_{mat}$ is $\ff{q^m}$-linear, this definition coincides with the usual definition of shortening on $\ff{q^m}$-linear codes. 
\end{defi}
This operation is interesting because it allows to decrease the number of $r_T$ variables in Modeling \ref{mod:RSL_minors_modeling} without altering the number of equations, which would be the case if we simply target a word of lower weight but without shortening. 

\subsection{Shortening $\mathcal{C}_{aug}$ as much as possible ($\delta = 0$).}\label{sec:zero}

A first idea is to look for a word of weight $r$ in a shortening of $\Cc_{aug}$ which contains roughly a unique word of this weight. Let $a \in \mathbb{N}$ be the unique integer such that $ar < N \leq (a+1)r$. From now on, we only consider  $N' = ar + 1$ errors. For $i \in \{1..N'\}$, we write $\ev_i = \Cm \times \Rm_i$, where the matrices $\Rm_1,\dots,\Rm_{N'}$ are random in $\ff{q}^{r \times n}$ and the matrix $\mat{C}$ is an $\ff{q}$-basis of $\mathcal{V}$. Thus, there exists roughly one linear combination of the $\ev_i$ of the form
\begin{equation*}
	\underline{\ev} = \Cm \times \begin{pmatrix} 0_{r \times a} & \widetilde{\Rm}  \end{pmatrix},
      \end{equation*}
where $\widetilde{\Rm}\in \ff{q}^{r \times (n-a)}$. A fortiori the error $\underline{\ev}$ lies in $\Cc_{aug}$, and its first $a$ coordinates are zero. In other words, the shortening $\mathcal{S}_{\lbrace 1..a \rbrace}(\mathcal{C}_{aug})$ contains about one word of weight $\leq r$. We use Modeling \ref{mod:RSL_minors_modeling} to attack this codeword, and the product $\Rm\trsp{\Hm}$ from the original system is replaced by $\widetilde{\Rm}\trsp{\widetilde{\Hm}}$, where the matrix $\widetilde{\mat{H}}$ consists of the last $n-a$ columns of $\Hm$ (note that we still have $\widetilde{\Hm}_{\any,\lbrace k+1..n\rbrace} = \Im_{n-k}$). The resulting system has roughly one solution. It consists of $m{n-k \choose r+1}$ equations but with only $N'$ variables $\lambda_i$ and ${n - a \choose r}$ variables $r_T$. Finally, the number of non-zero terms per equation is $N'\binom{k+1+r}{r}$.

\subsection{Looking for words of smaller weight in $\mathcal{C}_{aug}$ ($\delta > 0$).}\label{sec:un}

Let $d_{\Cc'}$ be the minimum distance of $\Cc'$ and  $\delta_{max} = r - d_{\Cc'}$. When $N$ is large enough, we have $\delta_{max} > 0$ and therefore there exist codewords of weight $w = r - \delta$ in $\Cc'$ for all $\delta \in \{1..\delta_{max}\}$. Once again, these codewords can be recovered by using Modeling \ref{mod:RSL_minors_modeling}. To estimate the number of solutions, we try to be more precise than the argument in \cite[C.1, Lemma 2]{GHPT17a} based on the rank Singleton bound and we use the following proposition. 
\begin{prop}\label{random}
	Let $r \in \mathbb{N}$ and $w \leq r$. Let $X_{\Cc',w}$  be the random variable counting the number of codewords of weight $w$ in $\Cc'$, where the randomness comes from the choice of a support $\mathcal{V}$ of dimension $r$ and of $N$ errors with support $\mathcal{V}$. The expectation and the variance of $X_{\Cc',w}$ are respectively given by
	\begin{equation*}
		\E{\left[ X_{\Cc',w} \right]} = \frac{\mathcal{S}_{w,r,n}}{q^{r \times n - N}} \text{ and }
	\end{equation*}
	\begin{equation*}
		\Var{\left[  X_{\Cc',w} \right]} = \mathcal{S}_{w,r,n} \times (q-1) \times \left( \frac{1}{q^{r \times n - N}} - \left(\frac{1}{q^{r \times n - N}}\right)^2 \right),
	\end{equation*}
	where $\mathcal{S}_{w,r,n}$ is the cardinality of the sphere of radius $w$ in $\mathbb{F}_q^{r \times n}$ for the rank metric.
\end{prop}
The proof can be found in Appendix \ref{sec:appendixB}. When $q$ is a constant, one obtains:
\begin{equation}\label{eq:mean}
	\E{\left[ X_{\Cc',w} \right]} =\Theta{(q^{w(n+r-w) - r \times n + N})} =  \Theta{(q^{N - (r-w)(n-w)})} \text{ and }
\end{equation}
\begin{equation}
	\Var{\left[ X_{\Cc',w} \right]} = \Theta{(q^{N + 1 - (r-w)(n-w)})} = \Theta{(q^{N - (r-w)(n-w)})}.
\end{equation}
Then, the code $\Cc'$ contains a word of weight $r-\delta$ with good probability whenever $N \geq \delta(n-r+\delta)$ using Equation \eqref{eq:mean}, and we look for such a codeword in the public code $\Cc_{aug}$. Also, when there are many of them, we proceed as in Section \ref{sec:zero} by shortening this code. For instance, if one has 
\begin{equation*}\label{eq:condi_delta_sup_0}
	N > \delta(n-r + \delta) + a \times (r-\delta),
\end{equation*}
we assume that there exists roughly one word of weight $\leq r-\delta$ in $\mathcal{S}_{\lbrace 1..a \rbrace}(\Cc')$. Therefore, the numbers of equations and monomials at bi-degree $(1,1)$ are now $m{n-k \choose r-\delta + 1}$ and $N'{n-a \choose r-\delta}$ respectively, where $N' = \delta(n-r + \delta) + a \times (r-\delta)$. The number of monomials per equation is $N'\binom{k+1+r-\delta}{r -\delta}$. In practice, we choose the value of $a$ which leads to the best complexity (see Table \ref{tab:rsl_long}).

\section{Solving the RSL Minors equations by linearization.}
\label{sec:attack2}
Now that we have restricted the number of solutions, we follow the approach from \cite{BBCGPSTV19} which consists in multiplying the bilinear equations by monomials in the $\lambda_i$'s and then solving by linearization at some bi-degree $(b,1)$ when there are enough equations compared to the number of monomials. In our case, the counting is much easier than in \cite{BBCGPSTV19} and we are able to determine with certainty the number of equations which are linearly independent over $\ff{q^m}$.  

\subsection{Number of independent equations for the system over $\ff{q^m}$.}\label{sec:grobner_rsl}

In this section, we focus on the initial system \eqref{eq:RSL_eq_fqm} whose equations are in $\ff{q^m}$. Our results rely on the following assumption. This assumption is very easy to check by linear algebra on the syndromes and was always verified in practice. 

\begin{assumption}\label{ass:syndrome}
	Let $\Sm =
	\begin{pmatrix}
		\trsp{\sv_1} & \dots & \trsp{\sv_N}
	\end{pmatrix} \in {\mathbb F_{q^m}}^{(n-k)\times N}$. We assume that
	$\Sm_{\{1..n-k-w\},\any}$ has rank $n-k-w$.
\end{assumption}

Under this assumption, we show that all the equations in system \eqref{eq:RSL_eq_fqm} are linearly independent over $\ff{q^m}$. The proof can be found in Appendix \ref{sec:appendixA}.

\begin{theorem}[Under Assumption \ref{ass:syndrome}]\label{distinct}
	The $\binom{n-k}{w+1}$ equations of system \eqref{eq:RSL_eq_fqm} are linearly independent over $\ff{q^m}$. 
\end{theorem}
 
As mentioned above, we are also interested in the number of independent equations over $\ff{q^m}$ at a higher bi-degree $(b,1)$ for $b \geq 2$. This number is not the maximal possible since linear relations between the augmented equations occur starting at $b=2$. However, this phenomenon is perfectly under control and Theorem \ref{indep} gives the exact number of independent equations at bi-degree $(b,1)$. Contrary to \cite{BBCGPSTV19}, this counting is still exact even when $b \geq w + 2$. 

\begin{theorem}[Under Assumption \ref{ass:syndrome}]\label{indep}
	For any $b\ge 1$, the $\mathbb F_{q^m}$-vector space
generated by the rows of the Macaulay matrix in bi-degree $(b,1)$ has dimension
\begin{equation}
	\mathcal{N}_{b} := \sum_{d=2}^{n-k-w+1} \binom{n-k-d}{w-1}\sum_{j=1}^{d-1}\binom{N-j+1+b-2}{b-1}.
\end{equation}
\end{theorem}

The proof can be found in Appendix \ref{sec:appendixB2}.

\subsection{Solving the RSL Minors equations by linearization.}\label{sec:(b,1)_rsl}

To obtain solutions over $\mathbb F_q$, one can expand each of the independent equations over $\ff{q^m}$ as $m$ equations over $\ff{q}$. We assume that linear relations do not occur after this process when there are less equations than the number of monomials in the resulting system. Recall also that this system has a unique solution. Therefore, the following Assumption \ref{ass:proj} gives the number of linearly independent equations at bi-degree $(b,1)$ at hand for any $b < q$. This assumption was somehow implicit in \cite{BBCGPSTV19} for the MaxMinors modeling, and it is also verified on our experiments in $\mathsf{magma}$. 
\begin{assumption}\label{ass:proj}
	For $b \geq 1$ and $b<q$, let $\mathcal{M}_b$ be the number of monomials at bi-degree $(b,1)$. Then, the number of linearly independent equations at bi-degree $(b,1)$ in the augmented system \eqref{eq:RSL_eq} is $m\mathcal{N}_b$ when $m\mathcal{N}_b < \mathcal{M}_b$, and $\mathcal{M}_b - 1$ otherwise, where $\mathcal{N}_b$ is defined as in Theorem \ref{indep}.
\end{assumption}
Combining Assumption \ref{ass:proj} and Theorem \ref{indep}, we obtain that one can solve by linearization  at bi-degree $(b,1)$ whenever $b < q$ and $m \mathcal{N}_{b} \geq \mathcal{M}_{b} - 1$, where $\mathcal{N}_{b}$ is defined above and $\mathcal{M}_{b} := {n \choose w}{N + b - 1 \choose b}$. However, we are mainly interested in the $q=2$ case, and due to the field equations we only have to consider squarefree monomials. The number of independent equations is now $m\mathcal{N}_{b}^{\ff{2}}$ when $m\mathcal{N}_{b}^{\ff{2}} < \mathcal{M}_{b}^{\ff{2}}$, where $\mathcal{M}_{b}^{\ff{2}} := {n \choose w}{N \choose b}$ and
\begin{align*}
	\mathcal{N}_{b}^{\ff{2}} & := \sum_{d=2}^{n-k-w+1}{n-k-d \choose w-1}\sum_{j=1}^{d-1}{N - j + 1 \choose b - 1}.
\end{align*}
In this case, it is favorable to consider all the equations up to bi-degree $(b,1)$ instead of those of exact bi-degree $(b,1)$. With $\mathcal{M}_{\leq b}^{\ff{2}} :=  \sum_{j=1}^b \mathcal{M}_{j}^{\ff{2}}$ and $\mathcal{N}_{\leq b}^{\ff{2}} :=  \sum_{j=1}^b \mathcal{N}_{j}^{\ff{2}}$, the condition to solve by linearization therefore reads
\begin{equation}\label{eq : lin_b_f2}
	m \mathcal{N}_{\leq b}^{\ff{2}} \geq \mathcal{M}_{\leq b}^{\ff{2}} - 1.
\end{equation}
The final linear system can be solved using the Strassen algorithm or the Wiedemann algorithm. If $b$ is the smallest positive integer such that \eqref{eq : lin_b_f2} holds, the complexities of solving this system are 
      \begin{equation}
	\label{complexity_q_minrank_S}
	\mathcal{O}\left(
	(\mathcal{N}_{\leq b}^{\ff{2}})(\mathcal{M}_{\leq b}^{\ff{2}})^{\omega - 1}
	\right)
\end{equation} and 
      \begin{equation}
	\label{complexity_q_minrank_W}
	\mathcal{O}\left(
	N\binom{k+1+w}{w}
	(\mathcal{M}_{\leq b}^{\ff{2}})^2
	\right) 
\end{equation}
field operations over $\ff{2}$ respectively, where $\omega$ is the linear algebra constant. Finally, one can use the hybrid approach \cite{BFP09} that performs exhaustive search in $\alpha_R$ variables $r_T$ and/or $\alpha_{\lambda}$ variables $\lambda_i$ in order to solve at a smaller bi-degree $(b,1)$,  and this strategy sometimes leads to better results (see Table \ref{tab:rsl_long}).

\section{Complexity of the attack on new Durandal parameters.}
\label{sec:tables}

We now present the best complexities obtained with our attack. In order to apply this attack to Durandal, we construct new parameters $(m,n,k,r,N)$ for the scheme by taking into account the constraints mentioned in \cite[§6.1]{ABGHZ19} but also the recent algebraic attacks from \cite{BBCGPSTV19}. The main ways to counteract these attacks are to increase the couple $(n,k)$ compared to $m$ or to increase the weight $r$, and the proposed parameters try to explore the two options. The missing parameters are chosen as follows. We always take $d=r$ and also $l' = 1$ in $N = (l + l')k$ as in \cite[§6.2]{ABGHZ19}. Apart from the key recovery attack, the most threatening attack against the scheme is the distinguishing attack on the $\text{PSSI}^{+}$ problem \cite[§4.1]{ABGHZ19}, which basically prevents from taking too small values for $m$ and $N$. Finally, for a given $4$-tuple $(m,n,k,r)$, we propose several values of $N$ to grasp how our attack behaves by increasing the number of errors.

In Table \ref{tab:rsl_long}, Column 2 refers to the distinguishing attack on $\text{PSSI}^{+}$. The cost corresponds to the advantage given in \cite[§4.1, Proposition $18$]{ABGHZ19}. The attacker is supposed to have access to $2^{64}$ signatures, so that the success probability must be $\leq 2^{-192}$ instead of $\leq 2^{-128}$. Column 3 refers to the RD attack from \cite{BBCGPSTV19} which is the best key recovery attack so far. The former combinatorial attack on RSL is much less efficient, so we do not even mention it. The rest of the table corresponds to our attack. We present the two ways to decrease the number of solutions described in Section \ref{sec:zero} ($``\delta = 0"$) and Section \ref{sec:un} ($``\delta > 0"$) and we give the value of $b$ to solve by linearization. Sometimes, the best strategy is the hybrid approach by fixing $\alpha_{\Rm}$ columns in $\Rm$ or $\alpha_{\lambda}$ variables $\lambda_i$. Also, the attack from Section \ref{sec:un} looks for a word of weight $w<r$ in $\Cc_{aug}$ and proceeds by shortening the matrices on $a$ columns. Therefore, the couple $(w,a)$ leading to the best complexity is also given. Finally, an underlined value represents an improvement upon the best RD attack, and a value in bold is a value below the 128-bit security level. 

\begin{table}[htb]
	\caption{Attack on 128-bit security parameter sets for Durandal. The missing parameters are chosen as in \cite{ABGHZ19}: we always take $d=r$ and also $l' = 1$ in $N = (l + l')k$. These choices only impact the $\text{PSSI}^{+}$ attack. Recall that the value in Column 2 must be $\geq 192$ assuming that the attacker has access to $2^{64}$ signatures. A starred value is obtained with the Wiedemann algorithm, otherwise the Strassen algorithm is used.}
	\begin{tabular}{|c|*{2}{@{}c@{}|}|@{~}c@{~}|c|@{}c@{}||c@{~}|c|c@{~}|c@{~}|@{}c@{}|}
		\hline
		
		$(m,n,k,r),N$  & $\text{PSSI}^{+}$ & Best RD & $\delta=0$ & $b$ & $(\alpha_{\Cm},\alpha_{\lambda})$ &  $\delta>0$ &  $b$ & $w = r - \delta$ & $a$ & $(\alpha_{\Cm},\alpha_{\lambda})$  \\\hline \hline
		
		$(277,358,179,7)$ &  &   & & & & & & & & \\
		
		$N = k(r-3)$ & 199 & 130 &   173 & 2 & (0,0) & $174^{*}$ & 3 & 6 & 60 & (0,0) \\
		
		$N = k(r-2)$ & 207 & 130 &   147 & 1 & (0,0) & \underline{\textbf{126}} & 1 & 5 & 37 & (0,2) \\
		
		$N = k(r-1)$  & 213 & 130 & 145 & 1 & (0,0) & \underline{\textbf{125}} & 1 & 5 & 19 & (0,1)  \\ 
		\hline

		$(281,242,121,8)$ &  &  &  & & & & & & & \\
		$N = k(r-2)$ & 193 & 159 &  $170$ & 2 & (0,0) &  $170^{*}$ & 3 & 7 & 70 & (0,0) \\
		$N = k(r-1)$ & 201  & 159 & $144$ & 1 & (0,0) &  $\underline{\textbf{128}}$ & 1 & 5 & 27 & (2,3) \\ 
		\hline
		
		$(293,254,127,8)$  &  &  &  & & & & & & &   \\ 
		$N = k(r-2)$ & 205 & 152  & $172$ & 2 & (0,0) & $172^{*}$ & 3 & 7 & 73 & (0,0) \\
		
		$N = k(r-1)$  & 213 & 152 & $145$ & 1 & (0,0) &    \underline{\textbf{125}} & 1 & 5 & 28 & (1,4)  \\ 
		\hline
		
		$(307,274,137,9)$  &  &  &  & & & & & & &   \\ 
		$N = k(r-2)$  & 199 & 251 & $187$ & 2 & (0,0) & $\underline{187}^{*}$ & 3 & 8 & 86 & (0,0)  \\ 
		$N = k(r-1)$  & 207 & 251 & $\underline{159}$ & 1 & (0,0) & $165^{*}$ & 2 & 8 & 103 & (0,0)  \\ 
		\hline
		
	\end{tabular}\label{tab:rsl_long}
\end{table}

 Our attack is very often more efficient than the best RD attack, and it is always the case when $N = (k-1)r$ in Table \ref{tab:rsl_long}. This improvement is not associated to a particular value of $N$ from which our attack will always be superior, but it is particularly obvious when the system can be solved at $b=1$. Note also that the progress is significant on the set of parameters with $r=9$, which suggests that our attack will be probably better for larger values of parameters as well. Finally, even if the cost of our attack is sometimes slightly below the $128$-bit security level, the effect on Durandal remains limited. This is mainly due to the fact that the attack on $\text{PSSI}^{+}$ is very powerful in a scenario in which the attacker has access to $2^{64}$ signatures.

\section{Conclusion.}
\label{sec:conclusion}
In this paper, we propose a new algebraic attack on RSL which clearly improves upon the previous attacks on this problem. As in \cite{BBCGPSTV19}, it relies on a bilinear modeling and avoids the use of generic Gröbner bases algorithms. However, the algebraic properties of our system allow a clearer analysis.

\begin{subappendices}
  \renewcommand{\thesection}{\Alph{section}}
  \setcounter{section}{0}
\section{Technical material.}\label{sec:appendixA}
This section contains the technical proofs of Lemma \ref{lemma:QJ}, Theorem \ref{distinct} and Theorem \ref{indep}. 
\paragraph{Proof of~\cref{lemma:QJ}.} We use the Cauchy-Binet formula to
express the determinant of a product of non-square matrices $A$ of
size $(w+1)\times n$ and $B$ of size $n \times (n-k)$,
  \begin{align*}
    \minor{AB}{} &= \sum_{\substack{T\subset \lbrace 1..n\rbrace\\\#T = w+1}}\minor{A}{\any,T}\minor{B}{T,\any}.
  \end{align*}
Then, we use the linearity of the determinant and Laplace expansion along first row of each left factor, to obtain, for any $J\subset\lbrace 1..n-k\rbrace$ of size $w+1$:
  \begin{align*}
    Q_J  &=\minor{\Delta_{\Hm}}{\any,J}=\sum_{\substack{T_0 \subset \lbrace 1..n \rbrace \\ \#T_0 = w+1}}\left\vert
    \begin{pmatrix}
      	\sum_{i=1}^{N} \lambda_{i}\yv_{i} \\
	\mat{C}
    \end{pmatrix}_{\any,T_0}\right\vert \left\vert \Hm \right \vert_{J,T_0}.
\notag{}\\
&= \sum_{i=1}^{N} \lambda_{i} \sum_{\substack{T \subset \lbrace 1..n \rbrace \\ \#T = w}} r_{T} \sum_{t \notin T} y_{i,t}(-1)^{1+Pos(t,T \cup \lbrace t \rbrace)}\left\vert \Hm \right \vert_{J,T \cup \lbrace t \rbrace}.
  \end{align*}
Without loss of generality, we assume from now on that  $\Hm_{\any,\lbrace k+1..n\rbrace} = \Im_{n-k}$. Then by Laplace expansion along columns in $\lbrace k+1..n\rbrace$, it is clear that if $T\cap\lbrace k+1..n\rbrace\not\subset(J+k)$, we have
  $ \left\vert \Hm \right \vert_{J,T \cup \lbrace t \rbrace}=0$ for
  any $t$ and the monomials involving $r_T$ do not appear in
  $Q_J$. There are at most $N\binom{k+w+1}w$ terms in $Q_J$, that can be written
  \begin{align}
Q_J    &= \sum_{i=1}^{N} \lambda_{i} \sum_{\substack{T \subset \lbrace 1..k \rbrace\cup (J+k) \\ \#T = w}} r_{T} \sum_{t \notin T} y_{i,t}(-1)^{1+Pos(t,T \cup \lbrace t \rbrace)}\left\vert \Hm \right \vert_{J,T \cup \lbrace t \rbrace}.\label{eq:QJ2}
  \end{align}

\paragraph{Proof of~\cref{distinct}.} We start by fixing a particular monomial ordering on $\mathcal{R}$ and we then provide, under~\cref{ass:syndrome}, a concrete linear transformation of
  the equations such that the resulting equations have distinct
  leading monomials. This will prove that they are linearly independent.
  
  Let $\prec$ be the grevlex monomial ordering on the variables
  $\lambda_i$ and $r_T$ such that
\begin{align*}
  r_{\lbrace t_1 < \dots < t_w \rbrace} \prec r_{\lbrace t'_1 < \dots < t'_w\rbrace} &\text{ iff }& t_i=t'_i \text{ for all } i < j \text{ and } t_j < t'_j,\\
r_T \prec  \lambda_N \prec \lambda_{N-1} \prec \dots \prec \lambda_{1} && \forall T \subset \lbrace 1..n\rbrace, \#T = w.
\end{align*}
This means that $\lambda_ir_T \prec \lambda_jr_{T'}$ iff $r_T\prec r_{T'}$ or $r_T = r_{T'}$ and $\lambda_i\prec\lambda_j$.

\begin{lemma}
For any $J\subset\lbrace 1..n-k\rbrace$ of size $w+1$, one can write
  \begin{align}
    Q_J &= \sum_{j\in J}\sum_{i=1}^N (-1)^{1+Pos(j,J)}s_{i,j}\lambda_i r_{(J\setminus \lbrace j\rbrace )+k} + \text{ (smaller terms wrt $\prec$)}.
  \end{align}
  where the smaller monomials are $\lambda_i r_T$ with
  $T\cap\lbrace 1..k\rbrace \neq \emptyset$, whereas the largest
  monomials are $\lambda_i r_T$ with
  $T\subset (J+k)\subset\lbrace k+1..n\rbrace$.
\end{lemma}
\begin{proof}
  We start from~\eqref{eq:QJ2}.  It is clear that the largest monomials
  are the $\lambda_j r_I$ with $I\subset \lbrace k+1..n\rbrace$, and
  they come from subsets $T\subset (J+k)$ of size $w$.

  Let $T\subset (J+k)$ and $\lbrace j \rbrace = J\setminus (T-k)$. We have
  \begin{align*}
    \left\vert \Hm\right\vert_{J,T\cup\{t\}} &=
    (-1)^{Pos(j,J)+Pos(t,T\cup\{t\})}h_{j,t}  &\text{ for } t\notin T\\
    h_{j,t}&=0 &\text{ for } t\in T
  \end{align*}
  (as $j\notin T$), so that the coefficient of $\lambda_i r_T$ in~\eqref{eq:QJ2} is
    \begin{align*}
    \sum_{t \notin T} y_{i,t}(-1)^{1+Pos(t,T \cup \lbrace t \rbrace)}\left\vert \Hm \right \vert_{J,T \cup \lbrace t \rbrace} &= \sum_{t\notin T} y_{i,t}h_{j,t}(-1)^{1+Pos(j,J)}\\
                                                                                                                                    &= (-1)^{1+Pos(j,J)}\sum_{t=1}^n y_{i,t}h_{j,t} \\&= (-1)^{1+Pos(j,J)} s_{i,j},
  \end{align*}
  and the last equality follows because $\sv_i = \yv_i\trsp{\Hm}$.
\end{proof}

We assume from now on that~\cref{ass:syndrome} is satisfied,
say the first $n-k-w$ rows of $\Sm =
\begin{pmatrix}
  \trsp{\sv_1} & \dots & \trsp{\sv_N}
\end{pmatrix}$ are linearly independent. 
This implies that, up to a permutation of the $\sv_i$'s, there
exist an invertible lower triangular matrix
$\Lm\in \mathbb F_{q^m}^{(n-k-w)\times (n-k-w)}$ and an
uppertriangular matrix $\Um
\in\mathbb F_{q^m}^{(n-k-w)\times N}$  such that all the entries of $\Um$ on its main diagonal are ones, and that 
$\Sm_{\lbrace 1..n-k-w\rbrace,\any} = \Lm\Um$.

We now construct the Macaulay matrix $\mathcal M$ associated to the
$Q_J$'s in bi-degree $(1,1)$, which is the matrix whose columns are labelled by 
the monomials $\lambda_i r_T$ sorted in descending order
w.r.t. $\prec$, whose rows correspond to the polynomials $Q_J$, and
whose entry in row $Q_J$ and column $\lambda_i r_T$ is the coefficient
of the monomial $\lambda_i r_T$ in the polynomial $Q_J$.

Let $I\subset\lbrace 1..n-k\rbrace$ of size $w$ and $i_1=\min(I)$. Consider 
the submatrix of the Macaulay matrix formed by the rows $Q_{\lbrace 1\rbrace \cup I},\dots, Q_{\lbrace i_1-1\rbrace \cup I}$. It has the shape
\begin{align*}
\mathcal M_I   &=\bordermatrix{
 & \dots & \lambda_1r_{I+k} & \dots&  \lambda_N r_{I+k}& \dots\cr
Q_{\lbrace 1\rbrace\cup I} & 0 & s_{1,1} & \dots& s_{N,1} & \dots\cr
Q_{\lbrace a\rbrace \cup I} & 0 & s_{1,a} &  & s_{N,a} &  \dots\cr
Q_{\lbrace i_1-1\rbrace \cup I}& 0 & s_{1,i_1-1} & & s_{N,i_1-1}  & \dots
} =
                                                                    \begin{pmatrix}
                                                                      \zerom & \Sm_{\lbrace 1..i_1-1\rbrace,\any} & \dots
                                                                    \end{pmatrix}.
  \end{align*}
Then we have
\begin{align*}
  (\Lm_{\lbrace 1..i_1-1\rbrace, \lbrace 1..i_1-1\rbrace})^{-1}  \mathcal M_I
  &=
    \begin{pmatrix}
    \zerom & \Um_{\lbrace 1..i_1-1\rbrace,\any} & \dots  
  \end{pmatrix}\\
  &= \bordermatrix{
 & \dots & \lambda_1r_{I+k} & \dots&\lambda_{i_1-1}r_{I+k}&  \dots\cr
& 0 & 1 &\dots& u_{i_1-1,1}& \dots\cr
& 0 & 0 &\ddots  & u_{i_1-1,a} &\dots\cr
& 0 & 0 &0 & 1 &  \dots
    }
\end{align*}
After applying those operations on all blocks, we have new equations $\tilde{Q}_{\lbrace i\rbrace\cup I}$ with leading monomials $\lambda_i r_{I+k}$.
Finally, any equation $Q_J$ has been transformed into an equation $\tilde{Q}_{J}$ with leading monomial $\lambda_{\min(J)} r_{(J\setminus \lbrace\min(J)\rbrace)+k}$, and they are all different, so that the equations are linearly independent.

\paragraph{Proof of~\cref{indep}.} \label{sec:appendixB2} We first
determine the number of linearly independent polynomials among the
polynomials $\lambda_j\tilde{Q}_{J}$. As the leading term of
$\tilde{Q}_J$ is
$\lambda_{\min(J)} r_{J\setminus\lbrace \min(J)\rbrace)+k}$, the
relations between the polynomials can only come from the pairs
$(\lambda_j\tilde{Q}_{\lbrace i \rbrace\cup I}, \lambda_i\tilde{Q}_{\lbrace j \rbrace\cup I})$ for all
$I\subset\lbrace 1..n-k\rbrace$ of size $w$ and all
$1\le i<j<\min(I)$. If we order the polynomials $\lambda_i\tilde{Q}_{\lbrace i\rbrace\cup I}$ such that
\begin{align*}
\lambda_i\tilde{Q}_{\lbrace j\rbrace\cup I}\prec \lambda_{i'}\tilde{Q}_{\lbrace j'\rbrace\cup I'}&\text{ iff}
                   \begin{cases}
                     {\lbrace j\rbrace\cup I}\prec_{lex}{\lbrace j'\rbrace\cup I'}&\text{ or}\\
                     {\lbrace j\rbrace\cup I}={\lbrace j'\rbrace\cup I'}\text{ and } i > i'
\end{cases}
\end{align*}
then it is clear that when we compute a row echelon form (without row pivoting) on the
Macaulay matrix in bi-degree $(2,1)$ for those polynomials in decreasing order, the only rows that can reduce to zero are the rows corresponding to the polynomials $\lambda_i\tilde{Q}_{\lbrace j\rbrace \cup I}$ with $I\subset\lbrace 1..n-k\rbrace$ of size $w$ and $1\le i < j < \min(I)$.
There are
$\sum_{i_1=1}^{n-k-w+1} \binom{i_1-1}2\binom{n-k-i_1}{w-1} =
\binom{n-k}{w+2}$ such polynomials.

On the other hand, the following~\cref{lemma:syzygies} provides the same number of
linearly independent relations between those polynomials,
which shows that the vector space
$\langle \lambda_j\tilde{Q}_J : 1\le j \le N, J\subset\lbrace
1..n-k\rbrace, \#J=w+1\rangle_{\mathbb F_{q^m}}$ is generated by the $\lbrace \lambda_j\tilde{Q}_J : \min(J) \le j \le N, J\subset\lbrace
1..n-k\rbrace, \#J=w+1\rbrace$ that
are linearly independent (they have distinct leading terms).

To conclude the proof of~\cref{indep} for any $b\ge 1$, it is readily seen that among the polynomials $\lambda_1^{\alpha_1}\cdots\lambda_N^{\alpha_N}\tilde{Q}_{\lbrace j_1\rbrace\cup I}$ for $\sum_i \alpha_i=b-1$, $j_1<\min(I)$, the ones with $\sum_{i=1}^{j_1-1}\alpha_i\ne 0$ reduce to zero (they are some multiple of a $\lambda_i\tilde{Q}_{J}$ with $i<\min(J)$ that reduces to zero), and the other polynomials have distinct leading terms
\begin{align*}
  \LT(\lambda_{j_1}^{\alpha_{j_1}}\cdots\lambda_N^{\alpha_N}\tilde{Q}_{\lbrace j_1\rbrace\cup I})
  &=\lambda_{j_1}^{\alpha_{j_1}}\cdots\lambda_N^{\alpha_N}\LT(\tilde{Q}_{\lbrace j_1\rbrace\cup I})
    =\lambda_{j_1}^{\alpha_{j_1+1}}\cdots\lambda_N^{\alpha_N}r_{I+k},
\end{align*}
The total number of such polynomials is:
\begin{align*}
  \sum_{i_1=2}^{n-k-w+1}\underbrace{\binom{n-k-i_1}{w-1}}_{\substack{ \text{number of sets } I\\ \text{with} \min(I)=i_1}}\sum_{j_1=1}^{i_1-1}\underbrace{\binom{N-j_1+1+b-2}{b-1}}_{\substack{\text{number of monomials in } \\ \lambda_{j_1},\dots,\lambda_N \text{ of degree } b-1.}}
\end{align*}

We are just left with the last lemma and its proof.
\begin{lemma}\label{lemma:syzygies}
	The following $\binom{n-k}{w+2}$ relations
	\begin{align*}
		\minor{
			\begin{pmatrix}
				\Delta_{\Hm}\\       \sum_{i=1}^N \lambda_i \sv_i
		\end{pmatrix}}{\any,K} &= 0 & \forall K\subset\lbrace 1..n-k\rbrace, \#K = w+2,
	\end{align*}
are relations
	between the $\lambda_jQ_J$'s (hence the $\lambda_j\tilde{Q}_J$'s), and
	under~\cref{ass:syndrome} they are linearly independent.
\end{lemma}
\begin{proof}
  Those minors are zero because the first and the last rows of the
  matrix are equal.  There are $\binom{n-k}{w+2}$ of them.  Laplace
  expansion along the last row gives, with
  $K=\lbrace k_1 < k_2 < \dots < k_{w+2}\rbrace$:
	\begin{align*}
		\sum_{i=1}^N \lambda_i \sum_{u=1}^{w+2} (-1)^{w+u}s_{i,k_u}Q_{K\setminus\{k_u\}}
		= \sum_{u=1}^{w+2} ((-1)^{w+u}\sum_{i=1}^N \lambda_i s_{i,k_u})Q_{K\setminus\{k_u\}}
	\end{align*}
	which corresponds to a syzygy
	\begin{align}
		{\mathcal G}^K = \begin{pmatrix}
			\underbrace{0}_{J\not \subset K}, \underbrace{(-1)^{w+u}\sum_{i=1}^N s_{i,k}\lambda_i}_{K\setminus J = \{k\}}
		\end{pmatrix}_{J\subset\{1..n-k\}, \#J=w+1}
	\end{align}
	If we order the $(Q_J)_J$ by decreasing lex ordering
	($Q_J\prec_{lex}Q_{J'}$ if $J\prec_{lex} J'$), then the first
	non-zero position of the syzygy $\mathcal G^K$ is the coefficient of
	the largest $Q_{K\setminus\{k_u\}}$, which is $Q_{K_1}$ with
	$K_1=K\setminus\lbrace k_{1}\rbrace$.
	The coefficient is:
	\begin{align*}
		(-1)^{w+1}\sum_{i=1}^N s_{i,k_1}\lambda_i
		= (-1)^{w+1}    \begin{pmatrix}
			s_{1,k_1}&\dots&s_{N,k_1}
		\end{pmatrix}\trsp{
			\begin{pmatrix}
				\lambda_1 & \dots & \lambda_N
		\end{pmatrix}}.
	\end{align*}
	This syzygie $\mathcal G^K$ shares the same leading position $Q_{K_1}$ with exactly the syzygies  $\mathcal G^{\lbrace j\rbrace \cup K_1}$ for  $1\le j< k_1$.
	If Assumption~\ref{ass:syndrome} is satisfied, then the coefficient in position $Q_{K_1}$ of 
	\begin{align*}
		(\Lm_{\{1..k_1\},\{1..k_1\}})^{-1}\begin{pmatrix}
			{\mathcal G}^{\lbrace 1\rbrace \cup K_1}\\
			\vdots\\
			{\mathcal G}^{\lbrace k_1-1\rbrace \cup K_1}\\
			{\mathcal G}^{K}
		\end{pmatrix}
	\end{align*}
	is
	\begin{align*}
		(-1)^{w+1}
		\begin{pmatrix}
			1 &\dots& u_{1,k_1}& \dots \\
			&\ddots  & \vdots &\dots \\
			0 & & 1 & \dots
		\end{pmatrix}
		\begin{pmatrix}
			\lambda_1\\\vdots\\\lambda_N
		\end{pmatrix}
	\end{align*}
	This shows that the syzygies
	$\mathcal G^{\lbrace j\rbrace\cup K_1}$,
	for $1\le j \le k_1$ are linearly independent. 
\end{proof}

\section{Number of low weight codewords in a random $\ff{q}$-linear code.}\label{sec:appendixB}
This section contains the proof of Proposition \ref{random} which provides formulae for the number of words of weight $\leq r$ in $\Cc'$ and, a fortiori, in $\Cc_{aug}$. Recall that $X_{\Cc',w}$ is the random variable which counts the number of codewords of weight $w$ in $\Cc'$. 
\paragraph{Proof of~\cref{random}.}
	Let $\Dc$ be the $\left[ r \times n , N \right]_{q}$-matrix code generated by the right factors $\Rm_i$, where $\ev_i := \Cm\Rm_i$ for $i \in \{1..N\}$. We assume that $\Dc$ is a random $\ff{q}$-linear code (see the end of Section \ref{sec:key_pair}). The matrix $\Cm$ has rank exactly equal to $r$, so that $X_{\Cc',w} = X_{\mathcal{D},w}$ for all $w \leq r$. For $\cv \in \mathbb{F}_q^{r \times n}$, we denote by $\mathsf{1}_{\cv \in \mathcal{D}}$ the random variable equal to $1$ if $\cv \in \mathcal{D}$ and $0$ otherwise, so that $X_{\mathcal{D},w} = \sum_{\omega(\cv)=w.}\mathsf{1}_{\cv \in \mathcal{D}}$. By linearity of expectation, one obtains:
	\begin{equation*}
		\E{\left[ X_{\mathcal{D},w} \right]} = \sum_{\omega(\cv)=w.}\E{\left[\mathsf{1}_{\cv \in \mathcal{C}}\right]} = \sum_{\omega(\cv)=w.}\mathcal{P}(\cv \in \mathcal{D}).
	\end{equation*}
	The probability that $\cv \in \mathcal{D}$ is the one to satisfy $r \times n - N$ independent parity-check equations of the form $\langle h , \cv \rangle = 0$, hence $\mathcal{P}(\cv \in \mathcal{D}) = \frac{1}{q^{r \times n - N}}$. The result follows by summing over all the codewords of weight $w$. For the variance, we start by computing the quantity
	\begin{equation*}
		\E{\left[ X_{\mathcal{D},w}^2 \right]} = \sum_{\omega(\cv_1)=w.}\sum_{\omega(\cv_2)=w.}\E{\left[\mathsf{1}_{\cv_1 \in \mathcal{D}} \mathsf{1}_{\cv_2 \in \mathcal{D}}\right]},
	\end{equation*} 
	and by definition we have $\E{\left[\mathsf{1}_{\cv_1 \in \mathcal{D}} \mathsf{1}_{\cv_2 \in \mathcal{D}}\right]} = \mathcal{P}(\cv_1 \in \mathcal{D},\cv_2 \in \mathcal{D})$. The code $\mathcal{D}$ being $\mathbb{F}_q$-linear, the events $\cv_1 \in \mathcal{D}$ and $\cv_2 \in \mathcal{D}$ are not independent when $\cv_2 \in \langle \cv_1 \rangle_{\mathbb{F}_{q}}$. In this case, one has
	\begin{equation*}
	\mathcal{P}_{\cv_2 \in \langle \cv_1 \rangle_{\mathbb{F}_{q}}}(\cv_1 \in \mathcal{D},\cv_2 \in \mathcal{D}) = \mathcal{P}(\cv_1 \in \mathcal{D}) =  \frac{1}{q^{r \times n - N}}.
	\end{equation*} 
	Therefore:
	\begin{align*}
		\E{\left[ X_{\mathcal{D},w}^2 \right]} & = \sum_{\omega(\cv_1)=w}\sum_{\substack{\cv_2 \in \langle \cv_1 \rangle_{\mathbb{F}_{q}} \\ \omega(\cv_2)=w}} \frac{1}{q^{r \times n - N}} + \sum_{\omega(\cv_1)=w}\sum_{\substack{\cv_2 \notin \langle \cv_1 \rangle_{\mathbb{F}_{q}} \\ \omega(\cv_2)=w}} \left(\frac{1}{q^{r \times n - N}}\right)^2 \\
		& = \mathcal{S}_{w,r,n}(q-1)\frac{1}{q^{r \times n - N}} + \mathcal{S}_{w,r,n}\left( \mathcal{S}_{w,r,n} - (q-1) \right) \left(\frac{1}{q^{r \times n - N}}\right)^2 \\
		& =	\E{\left[ X_{\mathcal{D},w} \right]}^2 + \mathcal{S}_{w,r,n} \times (q-1) \times \left( \frac{1}{q^{r \times n - N}} - \left(\frac{1}{q^{r \times n - N}}\right)^2 \right),
	\end{align*}
	hence the formula for the variance. 
\end{subappendices}

\end{document}